\input harvmac

\input amssym
\input epsf

\def\unit{\relax{\rm 1\kern-.26em I}}
\def\nada{\relax{\rm 0\kern-.30em l}}
\def\tilde{\widetilde}

%\draftmode

%\def\Omega{\rho,\sigma,\nu  }

%% MACROS
\noblackbox
\def\IL{\relax{\rm I\kern-.18em L}}
\def\IH{\relax{\rm I\kern-.18em H}}
\def\IR{\relax{\rm I\kern-.18em R}}
\def\IC{\relax\hbox{$\inbar\kern-.3em{\rm C}$}}
\def\IZ{\relax\ifmmode\mathchoice
{\hbox{\cmss Z\kern-.4em Z}}{\hbox{\cmss Z\kern-.4em Z}}
{\lower.9pt\hbox{\cmsss Z\kern-.4em Z}} {\lower1.2pt\hbox{\cmsss
Z\kern-.4em Z}}\else{\cmss Z\kern-.4em Z}\fi}
\def\CM {{\cal M}}

\def\CJ {{\cal J}}

\def\CO {{\cal O}}

%% MORE MACROS
\def\CM {{\cal M}}

\def\CO {{\cal O}}

\def\Tr{{\rm Tr}}

\font\manual=manfnt \def\dbend{\lower3.5pt\hbox{\manual\char127}}

\def\IZ{\relax\ifmmode\mathchoice
{\hbox{\cmss Z\kern-.4em Z}}{\hbox{\cmss Z\kern-.4em Z}}
{\lower.9pt\hbox{\cmsss Z\kern-.4em Z}} {\lower1.2pt\hbox{\cmsss
Z\kern-.4em Z}}\else{\cmss Z\kern-.4em Z}\fi}

\def\bar{\overline}

\def\rt2{\sqrt{2}}
\def\irt2{{1\over\sqrt{2}}}

%  \slashchar puts a slash through a character to represent contraction
%  with Dirac matrices. Use \not instead for negation of relations, and use
%  \hbar for hbar.
\def\slashchar#1{\setbox0=\hbox{$#1$}           % set a box for #1
\dimen0=\wd0                                 % and get its size
\setbox1=\hbox{/} \dimen1=\wd1               % get size of /
\ifdim\dimen0>\dimen1                        % #1 is bigger
   \rlap{\hbox to \dimen0{\hfil/\hfil}}      % so center / in box
   #1                                        % and print #1
\else                                        % / is bigger
   \rlap{\hbox to \dimen1{\hfil$#1$\hfil}}   % so center #1
   /                                         % and print /
\fi}

\def\foursqr#1#2{{\vcenter{\vbox{
 \hrule height.#2pt
 \hbox{\vrule width.#2pt height#1pt \kern#1pt
 \vrule width.#2pt}
 \hrule height.#2pt
 \hrule height.#2pt
 \hbox{\vrule width.#2pt height#1pt \kern#1pt
 \vrule width.#2pt}
 \hrule height.#2pt
     \hrule height.#2pt
 \hbox{\vrule width.#2pt height#1pt \kern#1pt
 \vrule width.#2pt}
 \hrule height.#2pt
     \hrule height.#2pt
 \hbox{\vrule width.#2pt height#1pt \kern#1pt
 \vrule width.#2pt}
 \hrule height.#2pt}}}}
\def\psqr#1#2{{\vcenter{\vbox{\hrule height.#2pt
 \hbox{\vrule width.#2pt height#1pt \kern#1pt
 \vrule width.#2pt}
 \hrule height.#2pt \hrule height.#2pt
 \hbox{\vrule width.#2pt height#1pt \kern#1pt
 \vrule width.#2pt}
 \hrule height.#2pt}}}}
\def\sqr#1#2{{\vcenter{\vbox{\hrule height.#2pt
 \hbox{\vrule width.#2pt height#1pt \kern#1pt
 \vrule width.#2pt}
 \hrule height.#2pt}}}}
\def\square{\mathchoice\sqr65\sqr65\sqr{2.1}3\sqr{1.5}3}

\def\figin{\epsfcheck\figin}\def\figins{\epsfcheck\figins}
\def\epsfcheck{\ifx\epsfbox\UnDeFiNeD
\message{(NO epsf.tex, FIGURES WILL BE IGNORED)}
\gdef\figin##1{\vskip2in}\gdef\figins##1{\hskip.5in}% blank space instead
\else\message{(FIGURES WILL BE INCLUDED)}%
\gdef\figin##1{##1}\gdef\figins##1{##1}\fi}
\def\DefWarn#1{}
\def\figinsert{\goodbreak\midinsert}
\def\ifig#1#2#3{\DefWarn#1\xdef#1{fig.~\the\figno}
\writedef{#1\leftbracket fig.\noexpand~\the\figno}%
\figinsert\figin{\centerline{#3}}\medskip\centerline{\vbox{\baselineskip12pt
\advance\hsize by -1truein\noindent\footnotefont{\bf
Fig.~\the\figno:\ } \it#2}}
\bigskip\endinsert\global\advance\figno by1}

%\MeadeWD
\lref\MeadeWD{ P.~Meade, N.~Seiberg and D.~Shih, ``General Gauge
Mediation,'' arXiv:0801.3278 [hep-ph].
%%CITATION = ARXIV:0801.3278;%%
}

%\MartinNS
\lref\MartinNS{
 S.~P.~Martin,
 ``A supersymmetry primer,''
 arXiv:hep-ph/9709356.
 %%CITATION = HEP-PH/9709356;%%
}

%\CheungES
\lref\CheungES{
 C.~Cheung, A.~L.~Fitzpatrick and D.~Shih,
 ``(Extra)Ordinary Gauge Mediation,''
 JHEP {\bf 0807}, 054 (2008)
 [arXiv:0710.3585 [hep-ph]].
 %%CITATION = JHEPA,0807,054;%%
}

%\GatesNR
\lref\GatesNR{
  S.~J.~Gates, M.~T.~Grisaru, M.~Rocek and W.~Siegel,
  ``Superspace, or one thousand and one lessons in supersymmetry,''
  Front.\ Phys.\  {\bf 58}, 1 (1983)
  [arXiv:hep-th/0108200].
  %%CITATION = FRPHA,58,1;%%
}

%\AffleckXZ
\lref\AffleckXZ{
  I.~Affleck, M.~Dine and N.~Seiberg,
  ``Dynamical Supersymmetry Breaking In Four-Dimensions And Its
  Phenomenological Implications,''
  Nucl.\ Phys.\  B {\bf 256}, 557 (1985).
  %%CITATION = NUPHA,B256,557;%%
}

%\DineYW
\lref\DineYW{
  M.~Dine and A.~E.~Nelson,
  ``Dynamical supersymmetry breaking at low-energies,''
  Phys.\ Rev.\  D {\bf 48}, 1277 (1993)
  [arXiv:hep-ph/9303230].
  %%CITATION = PHRVA,D48,1277;%%
}

%\DineVC
\lref\DineVC{
  M.~Dine, A.~E.~Nelson and Y.~Shirman,
  ``Low-Energy Dynamical Supersymmetry Breaking Simplified,''
  Phys.\ Rev.\  D {\bf 51}, 1362 (1995)
  [arXiv:hep-ph/9408384].
  %%CITATION = PHRVA,D51,1362;%%
}

%\DineAG
\lref\DineAG{
  M.~Dine, A.~E.~Nelson, Y.~Nir and Y.~Shirman,
  ``New tools for low-energy dynamical supersymmetry breaking,''
  Phys.\ Rev.\  D {\bf 53}, 2658 (1996)
  [arXiv:hep-ph/9507378].
  %%CITATION = PHRVA,D53,2658;%%
}

%\WittenKV
\lref\WittenKV{
  E.~Witten,
  ``Mass Hierarchies In Supersymmetric Theories,''
  Phys.\ Lett.\  B {\bf 105}, 267 (1981).
  %%CITATION = PHLTA,B105,267;%%
}

%\BanksMG
\lref\BanksMG{
  T.~Banks and V.~Kaplunovsky,
  ``Nosonomy Of An Upside Down Hierarchy Model. 1,''
  Nucl.\ Phys.\  B {\bf 211}, 529 (1983).
  %%CITATION = NUPHA,B211,529;%%
}
%\KaplunovskyYX
\lref\KaplunovskyYX{
  V.~Kaplunovsky,
  ``Nosonomy Of An Upside Down Hierarchy Model. 2,''
  Nucl.\ Phys.\  B {\bf 233}, 336 (1984).
  %%CITATION = NUPHA,B233,336;%%
}

%\DimopoulosGM
\lref\DimopoulosGM{
  S.~Dimopoulos and S.~Raby,
  ``Geometric Hierarchy,''
  Nucl.\ Phys.\  B {\bf 219}, 479 (1983).
  %%CITATION = NUPHA,B219,479;%%
}

%\DineGU
\lref\DineGU{
  M.~Dine and W.~Fischler,
  ``A Phenomenological Model Of Particle Physics Based On Supersymmetry,''
  Phys.\ Lett.\  B {\bf 110}, 227 (1982).
  %%CITATION = PHLTA,B110,227;%%
}

%\NappiHM
\lref\NappiHM{
  C.~R.~Nappi and B.~A.~Ovrut,
  ``Supersymmetric Extension Of The SU(3) X SU(2) X U(1) Model,''
  Phys.\ Lett.\  B {\bf 113}, 175 (1982).
  %%CITATION = PHLTA,B113,175;%%
}

%\DineZB
\lref\DineZB{
  M.~Dine and W.~Fischler,
  ``A Supersymmetric Gut,''
  Nucl.\ Phys.\  B {\bf 204}, 346 (1982).
  %%CITATION = NUPHA,B204,346;%%
}

%\AlvarezGaumeWY
\lref\AlvarezGaumeWY{
  L.~Alvarez-Gaume, M.~Claudson and M.~B.~Wise,
  ``Low-Energy Supersymmetry,''
  Nucl.\ Phys.\  B {\bf 207}, 96 (1982).
  %%CITATION = NUPHA,B207,96;%%
}

\lref\gmreview{
  G.~F.~Giudice and R.~Rattazzi,
  ``Theories with gauge-mediated supersymmetry breaking,''
  Phys.\ Rept.\  {\bf 322}, 419 (1999)
  [arXiv:hep-ph/9801271].
  %%CITATION = PRPLC,322,419;%%
}

%\SeibergQJ
\lref\SeibergQJ{
  N.~Seiberg, T.~Volansky and B.~Wecht,
  ``Semi-direct Gauge Mediation,''
  JHEP {\bf 0811}, 004 (2008)
  [arXiv:0809.4437 [hep-ph]].
  %%CITATION = JHEPA,0811,004;%%
}

\lref\dimgiud{
 S.~Dimopoulos and G.~F.~Giudice,
  ``Multi-messenger theories of gauge-mediated supersymmetry breaking,''
  Phys.\ Lett.\  B {\bf 393}, 72 (1997)
  [arXiv:hep-ph/9609344].
  %%CITATION = PHLTA,B393,72;%%
}

\lref\CarpenterWI{
  L.~M.~Carpenter, M.~Dine, G.~Festuccia and J.~D.~Mason,
  ``Implementing General Gauge Mediation,''
  arXiv:0805.2944 [hep-ph].
  %%CITATION = ARXIV:0805.2944;%%
}

\lref\NakayamaCF{
  Y.~Nakayama, M.~Taki, T.~Watari and T.~T.~Yanagida,
  ``Gauge mediation with D-term SUSY breaking,''
  Phys.\ Lett.\  B {\bf 655}, 58 (2007)
  [arXiv:0705.0865 [hep-ph]].
  %%CITATION = PHLTA,B655,58;%%
}

\lref\PoppitzXW{
  E.~Poppitz and S.~P.~Trivedi,
  ``Some remarks on gauge-mediated supersymmetry breaking,''
  Phys.\ Lett.\  B {\bf 401}, 38 (1997)
  [arXiv:hep-ph/9703246].
  %%CITATION = PHLTA,B401,38;%%
}

\lref\ArkaniHamedJV{
  N.~Arkani-Hamed, J.~March-Russell and H.~Murayama,
  ``Building models of gauge-mediated supersymmetry breaking without a
  messenger sector,''
  Nucl.\ Phys.\  B {\bf 509}, 3 (1998)
  [arXiv:hep-ph/9701286].
  %%CITATION = NUPHA,B509,3;%%
}

\lref\WessCP{
  J.~Wess and J.~Bagger,
  ``Supersymmetry and supergravity,''
%\href{http://www.slac.stanford.edu/spires/find/hep/www?irn=5426545}{SPIRES entry}
{\it  Princeton, USA: Univ. Pr. (1992) 259 p} }

\lref\AllanachNJ{
  B.~C.~Allanach {\it et al.},
  ``The Snowmass points and slopes: Benchmarks for SUSY searches,''
in {\it Proc. of the APS/DPF/DPB Summer Study on the Future of
Particle Physics (Snowmass 2001) } ed. N.~Graf
  [arXiv:hep-ph/0202233].
  %%CITATION = ECONF,C01063;%%
}

\lref\OoguriEZ{
  H.~Ooguri, Y.~Ookouchi, C.~S.~Park and J.~Song,
  ``Current Correlators for General Gauge Mediation,''
  Nucl.\ Phys.\  B {\bf 808}, 121 (2009)
  [arXiv:0806.4733 [hep-th]].
  %%CITATION = NUPHA,B808,121;%%
}

\lref\DistlerBT{
  J.~Distler and D.~Robbins,
  ``General F-Term Gauge Mediation,''
  arXiv:0807.2006 [hep-ph].
  %%CITATION = ARXIV:0807.2006;%%
}

\lref\IntriligatorFR{
  K.~A.~Intriligator and M.~Sudano,
  ``Comments on General Gauge Mediation,''
  JHEP {\bf 0811}, 008 (2008)
  [arXiv:0807.3942 [hep-ph]].
  %%CITATION = JHEPA,0811,008;%%
}

\lref\BenakliPG{
  K.~Benakli and M.~D.~Goodsell,
  ``Dirac Gauginos in General Gauge Mediation,''
  arXiv:0811.4409 [hep-ph].
  %%CITATION = ARXIV:0811.4409;%%
}

\lref\CarpenterHE{
  L.~M.~Carpenter,
  ``Surveying the Phenomenology of General Gauge Mediation,''
  arXiv:0812.2051 [hep-ph].
  %%CITATION = ARXIV:0812.2051;%%
}

%\DineGU
\lref\DineGU{
  M.~Dine and W.~Fischler,
  ``A Phenomenological Model Of Particle Physics Based On Supersymmetry,''
  Phys.\ Lett.\  B {\bf 110}, 227 (1982).
  %%CITATION = PHLTA,B110,227;%%
}

%\MartinZB
\lref\MartinZB{
  S.~P.~Martin,
  ``Generalized messengers of supersymmetry breaking and the sparticle mass
  spectrum,''
  Phys.\ Rev.\  D {\bf 55}, 3177 (1997)
  [arXiv:hep-ph/9608224].
  %%CITATION = PHRVA,D55,3177;%%
}

%\GiudiceNI
\lref\GiudiceNI{
  G.~F.~Giudice and R.~Rattazzi,
  ``Extracting Supersymmetry-Breaking Effects from Wave-Function
  Renormalization,''
  Nucl.\ Phys.\  B {\bf 511}, 25 (1998)
  [arXiv:hep-ph/9706540].
  %%CITATION = NUPHA,B511,25;%%
}

\lref\AllanachNJ{
  B.~C.~Allanach {\it et al.},
  %``The Snowmass points and slopes: Benchmarks for SUSY searches,''
in {\it Proc. of the APS/DPF/DPB Summer Study on the Future of
Particle Physics (Snowmass 2001) } ed. N.~Graf, {\it In the
Proceedings of APS / DPF / DPB Summer Study on the Future of
Particle Physics (Snowmass 2001), Snowmass, Colorado, 30 Jun - 21
Jul 2001, pp P125}
  [arXiv:hep-ph/0202233].
  %%CITATION = ECONF,C01063;%%
}

\lref\KSmuBmu{
  Z.~Komargodski and N.~Seiberg,
  ``$\mu$ and General Gauge Mediation,'' to appear.
}

\newbox\tmpbox\setbox\tmpbox\hbox{\abstractfont }
\Title{\vbox{\baselineskip12pt }} {\vbox{\centerline{ Exploring
General Gauge Mediation}}}
\smallskip
\centerline{Matthew Buican$^\dagger$, Patrick Meade$^*$, Nathan
Seiberg$^*$, and David Shih$^*$}
\smallskip
\bigskip
\centerline{$^\dagger${\it Department of Physics, Princeton
University, Princeton, NJ 08544 USA}} \centerline{$^*${\it School
of Natural Sciences, Institute for Advanced Study, Princeton, NJ
08540 USA}} \vskip 1cm

\noindent  We explore various aspects of General Gauge Mediation
(GGM). We present a reformulation of the correlation functions
used in GGM, and further elucidate their IR and UV properties.
Additionally we clarify the issue of UV sensitivity in the
calculation of the soft masses in the MSSM, highlighting the role
of the supertrace over the messenger spectrum. Finally, we present
weakly coupled messenger models which fully cover the parameter
space of GGM. These examples demonstrate that the full parameter
space of GGM is physical and realizable. Thus it should be
considered a valid basis for future phenomenological explorations
of gauge mediation.

\bigskip

\Date{December 2008}

\newsec{Introduction}

Low-energy supersymmetry, in its minimal incarnation as the MSSM,
is probably the most attractive candidate for physics beyond the
Standard Model, since it solves the hierarchy problem and predicts
gauge coupling unification.  However, the MSSM has one major
drawback, namely, its immense parameter space. Soft SUSY-breaking
introduces ${\cal O}(100)$ new parameters compared to the SM.
These parameters are highly constrained  by stringent experimental
limits on flavor-changing neutral currents and CP violation. A
conservative ansatz for the parameter space which is automatically
consistent with flavor and CP is known as ``soft SUSY-breaking
universality" (see \MartinNS\ for a nice review). Here there are
five flavor-diagonal sfermion masses, three real gaugino masses,
three flavor-diagonal $A$-terms, and three independent real Higgs
mass parameters, for a total of 14 real parameters in all. If one
accepts the hypothesis of universality, then the theoretical
challenge is to construct models of SUSY-breaking and mediation
that automatically produce universal patterns of soft parameters
without fine tuning.

Gauge mediation
\refs{\WittenKV\BanksMG\KaplunovskyYX\DimopoulosGM\DineGU\NappiHM
\DineZB\AlvarezGaumeWY\DineYW\DineVC-\DineAG}, or the idea that
SUSY-breaking is communicated to the MSSM via the SM gauge
interactions, is a promising partial solution to this
challenge.\foot{For a review of gauge mediation from both the
model building and phenomenological point of view see \gmreview.}
Since the gauge interactions are flavor blind, the soft masses
obtained through gauge mediation are automatically flavor
universal. However, the absence of CP phases is less automatic in
gauge mediation. Also, the Higgs $\mu$ and $B_\mu$ parameters are
not generated in pure gauge mediation, so one typically assumes
that additional interactions are present to produce these (for a
recent discussion of this see \KSmuBmu).

Recently in \MeadeWD, gauge mediation was given a general,
model-independent definition: {\it in the limit that the MSSM
gauge couplings $\alpha_i \rightarrow 0$, the theory decouples
into the MSSM and a separate hidden sector that breaks SUSY}. It
follows then that the SM gauge group must be part of a
weakly-gauged global symmetry $G$ of the hidden sector. By
studying a small set of current-current correlators of $G$, it was
shown that all the dependence of the soft masses on the hidden
sector could be encapsulated by three real parameters that
determine the sfermion masses, and three complex parameters that
determine the gaugino masses. This framework was called ``General
Gauge Mediation" (GGM) in \MeadeWD;  for more recent work on GGM,
see
\refs{\CarpenterWI\OoguriEZ\DistlerBT\IntriligatorFR\BenakliPG-\CarpenterHE}.
In this paper we will further develop several aspects of GGM and
explore its properties and its parameter space.

The definition of GGM must be augmented with several
phenomenological and consistency requirements, which we will now
review. First, the fact that the gaugino masses are complex in
general gauge mediation (GGM) implies that GGM does not solve the
SUSY CP problem. So additional mechanisms (such as an R-symmetry
as in \CheungES, or having the hidden sector be CP invariant) must
be invoked to explain why the gaugino masses are real.\foot{Of
course, one can have non-zero phases in this framework as long as
they are consistent with the experimental bounds. For convenience
though, we will only concentrate on CP invariant hidden sectors.}
For the rest of the paper, wherever it is relevant, we will {\it
assume} that such a mechanism is at work and only consider CP
invariant theories, so that the parameter space of GGM spans
${\Bbb R}^6$. With this assumption, the GGM parameter space
comprises a much smaller, but still sizeable subspace of the full
``universal" soft mass ansatz.

Additionally, as in \MeadeWD, we will impose a ${\Bbb Z}_2$
symmetry, called ``messenger parity," on our hidden sector.  In
the context of messengers this is typically defined as an
interchange symmetry of the messengers combined with $V\rightarrow
-V$ \refs{\DineGU,\dimgiud}.  More generally, messenger parity can
be defined  in terms of the gauge current and its supersymmetric
partners, without explicit reference to messengers \MeadeWD. This
symmetry does not have to be imposed, but it is typically a
phenomenological necessity: messenger parity prevents dangerous
hypercharge D-terms (which could lead to tachyonic sleptons) from
being generated in the hidden sector.

Messenger parity has various other consequences, including one on
the sum rules of GGM. The fact that the five flavor-diagonal
sfermion masses ($m_{Q}^2, m_{U}^2,m_{D}^2,m_{L}^2,m_{E}^2$) are
determined in terms of three real numbers implies that they must
satisfy two sum rules \MeadeWD:
 \eqn\massrell{\eqalign{
\Tr\,Y m^2\,\,&\propto\,\, m_{Q}^2-2
m_{U}^2+m_{D}^2-m_{L}^2+m_{E}^2 = 0 \cr \Tr\,(B-L)m^2
\,\,&\propto\,\, 2m_Q^2-m_U^2-m_D^2-2m_L^2+m_E^2=0.\cr }} These
sum rules are valid at the characteristic scale $M$ of the gauge
mediated model, and they are preserved by
the (one-loop) running of the soft masses in the MSSM. There could in
principle be violations to these sum rules
arising at higher order in the SM gauge couplings, coming from
$3$-point functions in the hidden sector. We will show in section
2 that in fact these threshold contributions satisfy the sum rules
if one imposes messenger parity on the hidden sector.
Additionally, the leading log contributions at all higher orders
also satisfy the sum rules. Therefore there are no contributions
at any relevant order in the hidden sector which would violate the
sum rules and they truly are predictions of GGM.

In \MeadeWD, it was shown that the GGM parameter space is the most
general that can be populated by models of gauge mediation.
However, this left open the important question of whether models
existed that could actually span this space.  For instance there
may have been additional relations or inequalities satisfied by
the parameters that were not manifest from the analysis of the
current-current correlators. Or it could have been that for some
regions of the GGM parameter space there was simply no field
theory that could populate it.  Indeed, a quick survey of existing
models of gauge mediation (e.g.\ the original models of ``minimal
gauge mediation"\refs{\DineYW,\DineVC}) would suggest that this
could be the case, as these models clearly do not cover the
parameter space. These models are based on a set of weakly coupled
``messengers," chiral superfields, $\Phi^i$, that transform under
a real representation of the SM gauge group and couple to a field
that has a SUSY breaking F-component. This can be expressed as
having a generic supersymmetric mass term for the messengers
\eqn\masswup{W=M_{ij}\Phi^i \Phi^j} and a SUSY-breaking mass term
of the form
 \eqn\Ftypeterms{V \supset f_{ij} \phi^i \phi^j + c.c.}
In \CarpenterWI\ it was shown that in the context of such models,
the right number (6) of parameters in GGM could be realized.
However, in their models the full space of GGM was not actually
spanned.

In this paper we further explore the model building possibilities
in the context of weakly coupled messengers and show that there
{\it are} models that span the GGM parameter space. This is
because there can be additional contributions to the MSSM soft
masses from gauge mediation in addition to those of the form
\Ftypeterms, namely ``diagonal-type"\refs{\PoppitzXW,\NakayamaCF}
messenger masses of the form \eqn\Dtypeterms{V \supset \xi_{ij}
\phi^i \phi^{\dagger j} } Such terms typically arise from D-term
breaking, but they can also arise from strong hidden sector
dynamics (such as in \SeibergQJ) where the distinction between
F-term and D-term breaking is not obvious.

Using both \Ftypeterms\ and \Dtypeterms, we demonstrate that there
exist weakly coupled messenger models which span the space of GGM.
Thus there can be no additional relations for the soft SUSY
breaking parameters beyond \massrell.

The outline of the paper is as follows. First, in section 2 we
present a reformulation of GGM that does not rely upon superspace
and that leads to extremely compact formulas for the gaugino and
sfermion soft masses. Using this formalism we will demonstrate
both the UV and IR finiteness of the soft masses in GGM. We will
then discuss in section 3 the dependence on the various mass
scales that can enter the correlation functions.  We will further
elaborate on the issues of UV sensitivity for SUSY breaking
parameters, clearing up some confusion in the existing literature
regarding the interpretation of a nonzero messenger supertrace.
Finally, in section 4 we present a simple explicit model involving
weakly-coupled messengers that spans the entire six-dimensional
parameter space of GGM. This model should be viewed merely as an
``existence proof" that the entire GGM parameter space can be
realized and that there are no additional hidden relations between
the parameters that are not obvious from the general formulation.
In light of this we believe that future phenomenological studies
of gauge mediation should not restrict themselves to the
parameterization of minimal gauge mediation (for example see
\AllanachNJ), but instead should explore the entire parameter
space of GGM. This should in principle open up new avenues for
possible experimental/phenomenological studies that have not yet
been explored (for recent work in this direction, see
\CarpenterHE). We finish by collecting a few technical results in
two appendices. In Appendix A we will review the role of the
supertrace in models with messenger fields.  We demonstrate that
certain classes of models always generate a particular sign for
the supertrace in an effective field theory. In appendix B we
collect some general results for the correlation functions of
models with arbitrary numbers of messengers.

\newsec{General Gauge Mediation: A New and Improved Formulation}

\subsec{Review and reformulation}

In this section we wish to review the basic features of GGM. Along
the way, we will reformulate and streamline various aspects of it.
This will lead to various new physical insights, including a
direct proof of the finiteness of the sfermion soft masses in GGM.

To begin, let us describe the setup. Consider a renormalizable
hidden sector\foot{We will consider non-UV-complete scenarios in
later sections.} which is characterized by the scale $M$ and where
supersymmetry is broken spontaneously. Suppose that this hidden
sector has a global symmetry group $G\supset G_{SM}=SU(3)\times
SU(2)\times U(1)$ that is weakly gauged. Suppose further that the
only coupling to the visible sector occurs through the SM gauge
interactions (so the hidden and visible sectors decouple in the
$g_{SM}\to 0$ limit). We will refer to this setup as general gauge
mediation, and we are interested in the visible-sector soft masses
that arise. As shown in \MeadeWD, all of the information in the
soft masses is encoded in two-point functions of the current
superfield of the symmetry group $G$.

To avoid writing all the gauge theory factors, we will assume for
simplicity that $G=U(1)$ in this subsection. Recall now the
definition of of the current superfield $\CJ$
 \eqn\CJdefD{D^2\CJ=0} which leads in components to
\eqn\CJdef{ \CJ = J+i\theta j-i\bar\theta\bar
j-\theta\sigma^\mu\bar\theta
j_\mu+{1\over2}\theta^2\bar\theta\bar\sigma^\mu \partial_\mu
j-{1\over2}\bar\theta^2 \theta \sigma^\mu\partial_\mu\bar j-
{1\over4}\theta^2\bar\theta^2 \square J } with $\partial^\mu
j_\mu=0$.

The use of superspace is not essential.  Without it, we can
replace the definition of the current superfield $\CJ$ \CJdefD\ as
follows.  We study the hermitian operator $J$ which satisfies
\eqn\QJ{ \{Q_\alpha,[Q_\beta, J]\} =0} where $Q_\alpha$ are the
supercharges, which satisfy the SUSY algebra
\eqn\susyalg{\{Q_\alpha, \bar Q_{\dot \alpha}\}=2
\sigma^\mu_{\alpha \dot\alpha} P_\mu.} Then, we can define
\eqn\compd{\eqalign{ &j_\alpha \equiv -i [Q_\alpha,J]\cr & j_{\dot
\alpha} \equiv i [\bar Q_{\dot \alpha},J]\cr & j_\mu \equiv
-{1\over 4}\bar\sigma_\mu^{\dot\alpha\alpha}\left(\{\bar
Q_{\dot\alpha},[Q_\alpha, J]\} - \{ Q_{\alpha},[\bar Q_{\dot
\alpha}, J]\}\right) ,}} and derive the current conservation by
applying two supercharges to this definition of $j_\mu$ and using
the SUSY algebra \susyalg.

The relation between the original presentation in superspace with
\CJdefD\ and this one is similar to the relation between the
definition of chiral superfields in terms of $\bar D \Phi=0$ and
the definition of chiral operators (the first component of $\Phi$)
as $[\bar Q , \phi]=0$.\foot{We will not pursue it here, but it
would be interesting to consider correlators of $J$'s defined by
\QJ\ along with any number of supercharges, in the case when SUSY
is unbroken. Perhaps there could be an interesting mathematical
structure analogous to operators in the chiral ring.} As we will
now show, \QJ\ proves to be extremely useful when computing
current-current correlation functions.

The correlators of interest are the nonzero current-current
two-point functions
 \eqn\currentspos{\eqalign{ & \langle
J(x) J(0)\rangle = {1\over x^4}C_0(x^2 M^2)\cr & \langle
j_{\alpha}(x) \bar j_{\dot\alpha}(0) \rangle
=-i\sigma^\mu_{\alpha\dot\alpha}\partial_\mu\left( {1\over
x^4}C_{1/2}(x^2 M^2)\right)\cr & \langle j_\mu(x) j_\nu(0)
\rangle= (\eta_{\mu\nu}\partial^2 -\partial_\mu
\partial_\nu)\left({1\over x^4}C_{1}(x^2 M^2)\right)\cr & \langle
j_\alpha(x)j_\beta(0)\rangle = \epsilon_{\alpha\beta} {1\over
x^5}B(x^2 M^2)\cr }} or in momentum space, \eqn\currents{\eqalign{
& \langle J(p) J(-p)\rangle =\tilde C_0(p^2/M^2)\cr & \langle
j_{\alpha}(p) \bar j_{\dot\alpha}(-p) \rangle
=-\sigma_{\alpha\dot\alpha}^\mu p_\mu \tilde C_{1/2} (p^2/M^2)\cr
& \langle j_\mu(p) j_\nu(-p) \rangle= -(p^2\eta_{\mu\nu} -p_\mu
p_\nu) \tilde C_1(p^2/M^2)\cr & \langle
j_\alpha(p)j_\beta(-p)\rangle = \epsilon_{\alpha\beta}M \tilde
B(p^2/M^2) }} where now a factor of $(2\pi)^4\delta^{(4)}(0)$ is
understood.

These two-point functions encode the mediation of SUSY breaking to
the MSSM gaugino and sfermion soft-masses at leading order in the
gauge coupling $g$. Specifically, the gaugino masses are given by
\eqn\mgaugino{ M_{gaugino}= g^2 M \tilde B(0). } while the
sfermion soft mass-squareds are given by \eqn\msfermion{
m_{sfermion}^2 = g^4 Y^2 A } where $Y$ is the $U(1)$ charge of the
sfermion and $A$ is the following linear combination of
correlators integrated over momentum: \eqn\Adef{\eqalign{ A
&\equiv -\int {d^4p\over (2\pi)^4}{1\over p^2}\Big(3 \tilde
C_1(p^2/M^2)-4 \tilde C_{1/2}(p^2/M^2)+ \tilde C_0(p^2/M^2)\Big)
\cr &=  -{M^2\over 16\pi^2}\int dy\,\Big(3 \tilde C_1(y) -4 \tilde
C_{1/2}(y)+\tilde C_0(y)\Big) }}

Using \QJ\ and \compd, one easily finds that formula for the
gaugino mass can be rewritten as \eqn\mgauginorew{ M_{gaugino} =
-{1\over4}g^2\int d^4x\,\langle Q^2(J(x)J(0))\rangle } where we
use the notation \eqn\notaQ{Q^2(\dots)=Q^\alpha
Q_\alpha(\dots)\equiv \{Q^\alpha,[Q_\alpha,(\dots)]\}.} Indeed,
according to \QJ\compd, $Q^2(J(x)J(0)) = 2[Q^\alpha ,J(x)]
[Q_\alpha ,J(0)]=-2j^\alpha(x)j_\alpha(0)$. Similar reasoning
shows that the action of four supercharges on $J(x)J(0)$ yields
\eqn\lincomboQiv{ \langle\bar Q^2(Q^2(J(x)J(0)))\rangle =
-8\partial^2(C_0(x)-4C_{1/2}(x)+3C_1(x)) } and so the formula for
the sfermion mass can be rewritten as \eqn\msfermionrew{\eqalign{
m_{sfermion}^2 &=-{1\over128\pi^2} g^4Y^2\int d^4x\,\log (x^2M^2)
\langle \bar Q^2 ( Q^2(J(x)J(0)))\rangle }} Note that the order of
the four supercharges is not essential -- a different ordering of
$Q$ and $\bar Q$ leads to terms that vanish after using the SUSY
algebra and momentum conservation. Note also that the scale $M$
appearing in \msfermionrew\ is arbitrary (i.e.\ the dependence on
$M$ drops out), since according to \lincomboQiv\ the integrand
$\langle \bar Q^2 ( Q^2(J(x)J(0)))\rangle$ is a total derivative.
(The short distance behavior of the correlator, to be discussed
below, guarantees that there is no surface term.)

Let us make some brief comments on the results \mgauginorew,
\msfermionrew. In \MeadeWD\ it was shown using the SUSY algebra
that when SUSY is unbroken, $B=0$ and $C_0=C_{1/2}=C_1$. Hence the
gaugino and sfermion masses vanish in the SUSY limit, as they
must. Writing the gaugino and sfermion masses as multiple
commutators, as we have done here, makes this fact obvious.

It is well known that when supersymmetry is broken at a scale $F$
and the dynamics is characterized by the scale $M \gg\sqrt F$, we
can effectively describe the soft terms in an expansion in $F
\over M^2$ using spurions.  Then the gaugino masses arise as an
F-term and the sfermion masses as a D-term.  The expressions
\mgauginorew\ and \msfermionrew\ generalize this result to the
more generic situation of $F\sim M^2$. The small $F \over M^2$
limit can be obtained by realizing that in \mgauginorew\ the two
$Q$s lead to one factor of $F$ and in \msfermionrew\ the four $Q$s
lead to $|F|^2$.

Another interesting feature of the formula \msfermionrew\ is that
all the information at large momentum is contained within the OPE
of $J$ with itself. This observation has immediate implications
about the convergence of the momentum integral in \Adef\ and
\msfermionrew. In \MeadeWD\ an indirect proof of the convergence
of these integrals was  given using the fact that otherwise there
would be no supersymmetric counterterm that could cancel a
divergence in this integral.  Here we can easily give a direct
proof which is intrinsic to the properties of the hidden sector.
The most singular term in the OPE $J(x)J(0)$ is associated with
the identity operator.  Since this is annihilated by the action of
the supercharges in \msfermionrew, to get a nonzero result we must
use an operator with $\Delta>0$.  Its coefficient is
$x^{-4+\Delta} $ and therefore the integral \msfermionrew\
converges at small $x$.

Finally, let us examine the low momentum behavior of the integral
in \Adef. We can exclude any zero-momentum divergences in these
integrals by invoking messenger parity $\CJ\to -\CJ$. On general
grounds, any such zero-momentum poles in the current two point
functions in \currents\ must be due to massless intermediate
one-particle states: \eqn\contJJ{ \langle \CJ(x)\CJ(0)\rangle =
\langle 0|\CJ(x)|\lambda\rangle\langle\lambda|\CJ(0)|0\rangle+...
} Assuming that the only massless particles in the spectrum are
due to spontaneously broken symmetries (bosonic or fermionic), and
that messenger parity commutes with all the symmetries of the
theory, it follows that the one-point functions on the RHS of
\contJJ\ must vanish. Therefore massless modes can never
contribute zero-momentum poles to the current two point function,
and the integral \Adef\ must always converge at $p=0$.

\subsec{ Generalization to the MSSM}

Finally, let us briefly generalize the discussion from our
$G=U(1)$ toy model to the MSSM, where $G=SU(3)\times SU(2)\times
U(1)$. We will label the gauge group factors $U(1)$, $SU(2)$ and
$SU(3)$ by $k=1,2,3$ respectively. Then are three complex numbers
$B_{k}\equiv\tilde B_k(0)$ and three real numbers $A_{k}$  which
determine the gaugino and sfermion soft masses. They are defined
as above, using the current supermultiplet of the respective gauge
group. The soft masses are given to leading order in the $\alpha$
by \eqn\softmassesMSSM{
 M_k = g_k^2 M B_k,\qquad m_{f}^2 = \sum_{k=1}^3 g_k^4 c_2(f,k) A_k
 }
 $f=Q,U,D,L,E$ labels the matter representations of the MSSM, and $c_2(f,k)$ is the quadratic Casimir of $f$ with respect to the gauge group $k$.

Since the five sfermion masses are determined by three real
numbers, they must satisfy two sum rules. These take the form
\MeadeWD: \eqn\massrelltwo{\eqalign{ & m_{Q}^2-2
m_{U}^2+m_{D}^2-m_{L}^2+m_{E}^2 = 0 \cr &
2m_Q^2-m_U^2-m_D^2-2m_L^2+m_E^2=0.\cr }} {}From \softmassesMSSM,
it is clear that these sum rules are valid at $\CO(\alpha^2)$.
However, we can further demonstrate that they are valid at
$\CO(\alpha^3)$ and to leading-log order for any $\alpha$, meaning
that the sum rules must be satisfied to very high accuracy.

First, it was already shown in \MeadeWD\ that the sum rules are
preserved by the MSSM RGEs (neglecting contributions from the
Higgs sector proportional to the Yukawa interactions). This takes
care of the leading-log corrections. Second, we can consider the
$\CO(\alpha^3)$ corrections coming from the hidden sector. These
arise from various current three-point functions in the hidden
sector. It is easy to see that gauge invariance allows only five
three-point functions: $SU(3)^3$, $SU(2)^3$, $U(1)^3$,
$SU(3)^2U(1)$, $SU(2)^2U(1)$. If one imposes messenger parity
(which sends $V_Y\to -V_Y$), this eliminates the mixed three-point
functions and the $U(1)^3$, leaving us with only the $SU(3)^3$ and
$SU(2)^3$ three point functions. These represent additional
contributions to the parameters $A_2$ and $A_3$. Their presence
does not spoil the sum rules, which only rely on the fact that
there are three $A$'s and not that they only receive contributions
at a given order in $\alpha$.

\newsec{Sensitivity to UV physics}

\subsec{General remarks}

In the previous section, we restricted our analysis to
renormalizable, UV-complete hidden sectors. However, it is often
the case that our understanding of the hidden sector is
incomplete, that we have only an effective description of it at
low energies. In this section we would like to make some general
comments about the dependence of the MSSM soft-breaking terms on
unknown UV physics. This will have immediate applications in the
next section, when we wish to use incomplete messenger-spurion
models of gauge mediation to cover the parameter space of GGM.
With our understanding of the (in)sensitivity of gauge mediation
to UV physics, we will be sure that the models we study in the
next section are indeed calculating correctly the MSSM soft
masses.

We will begin with a more abstract discussion of UV sensitivity in
a theory with spontaneously broken SUSY. Then in the next
subsection we will give an example to illustrate some of our
general comments.  The reader may find it useful to reread the
general discussion after having gone through the example
calculation in the next subsection.

Consider a hidden sector consisting of an effective field theory
valid below a UV cutoff scale $\Lambda$ (which could be e.g.\ the
Planck scale, or some UV scale), with SUSY spontaneously broken at
a scale $\sqrt{F}$. As long as $\sqrt F \ll \Lambda $, all the
soft terms are calculable in terms of the effective theory.  The
reason is that at energies much larger than $\sqrt F$
supersymmetry is restored and all the supersymmetry breaking
contributions arise at energies of order $\sqrt F$ or smaller.

Now suppose the hidden sector is a messenger model of gauge
mediation. Such models are weakly coupled truncations of a more
complete theory valid above the scale $\Lambda$. They are fully
specified by the set of messenger quantum numbers and the set of
messenger masses given in \masswup, \Ftypeterms, \Dtypeterms. In
this scheme, the soft parameters are calculable in terms of the
messenger mass matrices. Let us denote the scale of the messenger
sector by $M$. Clearly, when we study these models, we are
implicitly taking the limit $\Lambda \to \infty$ with $M$ fixed.

Typically one considers the messenger scale $M$ and the
SUSY-breaking scale $\sqrt{F}$ to be of the same order. In this
case there is no problem and the soft terms are indeed
unambiguously calculable, insensitive to the physics above the UV
cutoff $\Lambda$.  However, it is often the case that the
messengers at the scale $M$ receive supersymmetry breaking mass
splittings which are much smaller than $F\over M$.  Then, we might
want to reconsider the $\Lambda \to \infty$ limit in such a way
that the messenger mass splittings are kept finite.

For example, imagine that these mass splittings are or order $F
\over \Lambda$. Then, the proper decoupling limit is $\Lambda,
\sqrt F \to \infty$ with fixed $F \over \Lambda$ and $M$.  In this
case the soft-breaking terms may not be calculable.  A simple way
to see that is to add to the theory additional messengers with
mass of order $\Lambda$ and supersymmetry breaking mass splittings
of order $F\over\Lambda$. These messengers contribute to gaugino
masses and sfermion mass-squareds additional terms of order ${F
\over \Lambda}$ and $({F \over \Lambda})^2$ respectively. We can
view these additional contributions as {\it finite local
counterterms} for gaugino masses and sfermion masses which are
determined by the details of the high energy theory.

{}From the point of view of the effective theory, such
counterterms are ambiguous, controlled by the choice of UV
completion above the scale $\Lambda$. It is important to note,
however, that any such ambiguity must necessarily arise {\it only
at leading order} in the SUSY breaking parameter $F$, since
higher-order contributions from the UV states are necessarily
suppressed by additional powers of ${F\over\Lambda^2}$ (which goes
to zero as $\Lambda\to\infty$).

The sensitivity to the UV is particularly dramatic when the
supertrace of the messenger spectrum is nonzero
\refs{\PoppitzXW,\ArkaniHamedJV}. In this case the necessary
counterterms include a logarithmically divergent sfermion mass.
(See Appendix B for an explicit proof of this fact.) We stress
that this divergence is a symptom of the problem, but the problem
might arise even if the supertrace vanishes.

We conclude by roughly summarizing the foregoing discussion: if
the messenger splittings are parametrically smaller than $F/M$,
the soft-breaking terms in the MSSM are not calculable without
further UV input.

\subsec{Example}

Let us now illustrate these general points with a simple example.
To that end, consider the messenger theory with superpotential
\eqn\Weffi{ W_{\rm eff}=M\phi_1\tilde\phi_1 } and K\"ahler
potential \eqn\Keffi{ K_{\rm
eff}=|X|^2+|\tilde\phi_1|^2+\Big(1+\Big|{X\over
\Lambda}\Big|^2+...\Big)|\phi_1|^2 } where the ellipsis contains
higher dimensional operators and $X$ is a SUSY breaking field with
\eqn\X{ \langle X\rangle=M'+\theta^2 F } It will be convenient to
introduce the following notation: \eqn\xynotation{ x \equiv
{M'\over \Lambda},\qquad y = {F\over M\Lambda} } As described
above, we consider the limit $\Lambda\to\infty$ with $x$ and $y$
and the low energy mass parameter $M$ held fixed.

By  the general arguments above, we expect that the soft
parameters computed in this effective theory are sensitive to
large corrections from states at the scale $\Lambda$ where the
description of the physics given by \Weffi\ and \Keffi\ breaks
down. Moreover, we expect that such corrections only enter in at
leading order in the SUSY-breaking parameter $F$. We will now
explicitly show that this is indeed the case.

Using our messenger GGM formalism developed in Appendix B, or
equivalently in this case using the explicit formulas from
\PoppitzXW, we find the low energy soft parameters to be
\eqn\BIR{\eqalign{ B_{\rm eff}&={Mx\over
48\pi^2(1+x^2)^2}\left(6(1+x^2) y+(2+x^2) y^3\right)+\CO(y^5) }}
and \eqn\AIR{ A_{\rm
eff}={M^2\over64\pi^4(1+x^2)^2}\left(\Big(\log\Big({\Lambda_{\rm
cutoff}^2\over M^2}\Big)-2+x^2+2\log(1+x^2)\Big)
y^2+{x^2(6+x^2)\over36(1+x^2)} y^4\right)+\CO(y^6) .} Note that
while $B_{\rm eff}$ is finite, $A_{\rm eff}$ is logarithmically
divergent with the UV cutoff $\Lambda_{\rm cutoff}$.  The
appearance of this divergence which multiplies the supertrace in
the low energy effective theory \eqn\effstr{ {\rm STr}\CM^2_{\rm
IR}=-{2M^2y^2\over(1+x^2)^2} } reminds us that our theory must be
UV completed. Note, however, that even though the gaugino mass
parameter is finite, it too will be sensitive to the UV physics as
we will see below.

We can regulate the divergence in \AIR\ by embedding the IR theory
in a renormalizable UV theory with the following superpotential
\eqn\testtheo{
W=X\phi_1\tilde\phi_2+M\phi_1\tilde\phi_1+\Lambda\phi_2\tilde\phi_2
} and a canonical K\"ahler potential.\foot{Some authors (see
e.g.\PoppitzXW) regularize the theory using dimensional reduction
with ``$\epsilon$-scalars.''  We prefer to replace the unphysical
$\epsilon $-scalars with physical heavy fields as in \testtheo.}
Integrating out the heavy fields (with mass $\Lambda$) $\phi_2$,
$\tilde\phi_2$, we readily derive the effective low energy
Lagrangian \Weffi, \Keffi.\foot{In this regularization, we see
that the negative sign of the supertrace in \effstr\ corresponds
precisely to what we expect from the general results on
integrating out massive chiral matter in Appendix A.}

The contribution of the messengers in our full theory \testtheo\
to the soft SUSY breaking masses in the MSSM is manifestly finite.
Let's compare it to the calculation in the low energy theory \BIR,
\AIR.

Again, using our messenger GGM formulas we find the following soft
parameters \eqn\BUV{\eqalign{ B_{\rm
full}&={Mx\over48\pi^2(1+x^2)^2}(2+x^2) y^3+\CO(y^5) }} and
\eqn\AUV{ A_{\rm
full}={M^2\over64\pi^4(1+x^2)^2}\left(\Big(\log\Big({\Lambda^2\over
M^2}\Big)+2x^2+2\log(1+x^2)\Big) y^2+{x^2(6+x^2)\over36(1+x^2)}
y^4\right)+\CO(y^6) }

We see that $B_{\rm eff}$ and $B_{\rm full}$
differ {\it only} at leading order in $y$, with the counterterm
given by\foot{One can check that the full expressions for both $B$
and $A$ in the effective and the full theories agree at {\it all}
higher orders in $y$ and not just at the next-to-leading order we
have written down in our expressions above.}
\eqn\Bdelta{ \delta
B={M\over8\pi^2}\Big({x\over1+x^2}\Big)y }
For the particular UV
definition we have chosen, we can understand this term as arising
from the rescaling anomaly in the recanonicalization of the IR
K\"ahler potential. Notice, however, that if we had added
messengers to the UV theory that did not couple to the light
messengers, they would have also contributed at order $y$ to the
counterterm in \Bdelta. These contributions cannot be captured by
the rescaling anomaly.

Similarly, the difference between
$A_{\rm full}$ and $A_{\rm eff}$ is also only at leading order in the SUSY
breaking. However, here it includes an infinite counterterm:
\eqn\Adelta{ \delta
A={M^2\over64\pi^4(1+x^2)^2}(\log(\Lambda^2/\Lambda_{\rm
cutoff}^2)+x^2+\log(1+x^2))y^2\ . }
Again, adding
messengers in the UV decoupled from the IR has the effect of generating additional
corrections at leading order in the SUSY breaking.

With a sharp set of criteria for defining calculable gauge
mediation models in hand, we will now explore the covering of the
GGM parameter space in the next section. In particular, when using
messenger models we will specialize to the case of vanishing
supertrace and ${F\over\Lambda}\to0$.

\newsec{Covering the General Gauge Mediation Parameter Space}

\subsec{The general setup}

In this section we will demonstrate, using a general model with
messengers, that the entire parameter space of GGM can be covered
by a calculable weakly coupled field theory.

Consider a theory with $N$ chiral messengers $\Phi^i$,
$\tilde\Phi^i$, $i=1,\dots,N$ transforming in some vector-like
representation ${\bf R}\oplus {\bf \bar  R}$ of a gauge group $G$
(which will later be identified with the SM gauge group). The
messenger spectrum determines the GGM soft masses, so we will
focus on that. The most general messenger spectrum is of the form
\eqn\Vgenspec{ V_{\rm mass\, terms} = (\tilde\psi^T
\CM_F\psi+c.c.) +\pmatrix{\phi \cr \tilde\phi^*}^\dagger
\CM_B^2\pmatrix{\phi\cr\tilde\phi^*} } with \eqn\MFMBdef{ \CM_B^2
\equiv \pmatrix{ \CM_F^\dagger \CM_F+\xi & F \cr F^\dagger &
\CM_F\CM_F^\dagger+\tilde\xi} } Here $\CM_F$, $\xi$, $\tilde\xi$
and $F$ are all $N\times N$ matrices. We take $\CM_F$ to be
diagonal with real, positive entries without loss of generality.
$\xi$ and $\tilde\xi$ are Hermitian; and $F$ is complex. The
off-diagonal parameters $F$ can arise from ``F-term breaking''
e.g.\ from a superpotential coupling to spurion field.  The
diagonal parameters $\xi$ can arise from ``D-term breaking'' e.g.\
from FI-U(1) terms. More generally, the general spectrum shown in
\Vgenspec\ can arise from complicated non-Abelian dynamics such as
in \SeibergQJ.

We will impose the following restrictions on the messenger
spectrum, motivated by phenomenology and overall consistency:

\item{1.} In order to avoid the SUSY CP problem, we require all
the mass parameters to be real
 \eqn\CPsym{\xi=\xi^*, \qquad \tilde\xi=\tilde\xi^*,\qquad F=F^*\ .}

\item{2.} In order to guarantee that no dangerous FI-term for
hypercharge is generated, we impose invariance under messenger
parity \refs{\DineGU,\dimgiud}\foot{Actually, the authors of
\dimgiud\ considered another action for this symmetry  which maps
chiral superfields to anti-chiral superfields.  Such a symmetry
does not commute with the Lorentz symmetry.  However, if we also
impose CP symmetry, our choice is equivalent to theirs.}
 \eqn\messpar{\Phi^i \leftrightarrow \tilde \Phi^i.}
This restricts the parameters to satisfy \eqn\messparcons{
\xi=\tilde\xi\, , \qquad  F=F^T\ . }

\item{3.} Since we want our theory to be calculable and
insensitive to UV physics, we require vanishing messenger
mass-squared supertrace.  This translates to
    \eqn\strzero{ \Tr\,\xi=0}

\item{4.} In the case where $G=SU(3)\times SU(2)\times U(1)$, we
want the gauge couplings to unify. This restricts the messengers
to be in complete $SU(5)$ representations.  Furthermore, we limit
the number of representations such that the theory remains
perturbative.

\item{5.} The messengers must be non-tachyonic for consistency of
the model. So this puts upper limits on the magnitudes of the
entries in $\xi$ and $F$.

\medskip

Finally, we note that if the messengers are in a reducible
representation \eqn\redrep{ {\bf R}=\bigoplus_R (n_R\times R) }
then the messenger mass matrices must be block-diagonal. Each
block couples the messengers with the same $R$. Consequently, all
of the statements above hold for each $R$ separately, and the
leading-order in $\alpha$ contributions from each $R$ to the soft
masses are additive.

\subsec{Covering the GGM parameter space of a toy $U(1)$ visible
sector}

In this subsection we will consider a simplified theory with only
$G=U(1)$ symmetry and messengers with charges $\pm 1$.  This
example is instructive because the detailed representation theory
of the messengers does not play an important role in this case.
It will also be useful in the next subsection when we consider the
full $G=SU(3)\times SU(2)\times U(1)$ case.

Here there is only one $A$ parameter and only one $B$ parameter
and covering the parameter space means finding a theory that
covers the range \eqn\covercondtwo{ \kappa={A\over
|B|^2}\in(0,\infty). } Notice that $\kappa\to0$ corresponds to the
limit of either a very massive gaugino or vanishing sfermion mass,
while $\kappa\to\infty$ corresponds to either a very massive
scalar or vanishing gaugino mass.

Let us first ask if we can cover \covercondtwo\ with a single
messenger pair and, at the same time, obey the microscopic
constraints on our messenger sector described in the previous
subsection. To answer this question, note that the most general
single messenger model allowed by messenger parity and vanishing
supertrace is of the form \eqn\fermandbosmass{ \CM_F=M, \qquad
\CM_B^2=\pmatrix{M^2 & F \cr F& M^2}. } i.e.\ only minimal gauge
mediation is allowed. This model has two parameters, $M$ and $F$,
and spans a two-dimensional subspace of the full $A$ and $B$
parameter space. However, an explicit calculation shows \MartinZB\
that this subspace is not the full GGM parameter space and that in
fact \eqn\MGMGbd{ \kappa\in(.37,1) } where the upper bound for
$\kappa$ is obtained in the limit of small SUSY breaking and the
lower bound arises because the messengers cannot be tachyonic.

Next, we try a system with two messengers. Since we are only
interested in giving an existence proof of \covercondtwo, we will
not consider the most general possible two-messenger mass matrix
satisfying the conditions above.  Instead, we consider the
following special mass matrix \eqn\mfermion{\CM_F=\pmatrix{M_1 & 0
\cr 0& M_2}} and \eqn\mboson{\CM_B^2= \pmatrix{ M_1^2 + D & 0 &
F_1 & 0\cr 0 & M_2^2 -D & 0 & F_2\cr F_1 & 0& M_1^2 + D & 0 \cr 0
& F_2& 0 & M_2^2 - D} .} This model could arise, e.g.\ from a
simple MGM-like setup with the messengers charged under an
additional $U(1)'$ gauge group with a nonzero FI D-term.

With the added assumption
 \eqn\range{F_1, F_2, D \ll
M_{1,2}^2 \ .} we can use the techniques of wavefunction
renormalization \refs{\GiudiceNI,\CheungES}\ to compute the $A$
and $B$ parameters \eqn\Bli{ B={1\over8\pi^2}\Big({F_1\over M_1} +
{F_2\over M_2}\Big)+\CO(F^3, D F) } and \eqn\Aa{\eqalign{ &A=A_F+
A_{\xi} \cr &A_F={1\over64\pi^4}\Big({F_1^2\over M_1^2} +
{F_2^2\over M_2^2}\Big)+\CO(F^4, D F^2) \cr &A_{\xi}=
{D\over32\pi^4}\log(M_1^2/M_2^2)+\CO(D F^2). }} {}From these
expressions, it is straightforward to see that this example in
fact covers the range \eqn\coverresult{\kappa\in(-\infty, \infty)
.} First, for $D=0$ we can set ${F_1 \over M_1} \approx -{F_2\over
M_2}$ such that $B$ is very small while $A$ is finite.  This leads
to arbitrarily large $|\kappa|$.  However, setting $D=0$ prevents
us from making $|\kappa|$ arbitrarily small.  For that, we use
nonzero $D$ to set
 \eqn\parrange{A_\xi <0}
such that $A=A_F + A_\xi$ is arbitrarily small with fixed $B$.

We conclude that this example covers the full parameter space of
GGM for a $U(1)$ visible sector.

\subsec{Covering the MSSM GGM parameter space}

Let us now generalize the discussion of the previous section to
the physically relevant case of $G=SU(3)\times SU(2)\times U(1)$.
We will see that, when properly analyzed, this case reduces to the
$U(1)$ case considered in the previous subsection.

We would like to find weakly-coupled messenger theories that cover
the full GGM parameter space of the MSSM, namely the six
parameters $A_k$, $B_k \in {\Bbb R}^+$, where $k=1,2,3$ labels
$U(1)$, $SU(2)$ and $SU(3)$,  respectively.   A first analysis of
this subject was presented by Carpenter, Dine, Festuccia and Mason
in \CarpenterWI.  We will extend their analysis, by demanding not
only the right number of parameters, but that the entire parameter
space can be covered.

As noted above around equation \redrep, the messenger mass
matrices are block diagonal with respect to different irreps $R$,
and the contribution from messengers of different irreps are
additive. It follows then that
 \eqn\addcon{A_k=\sum_R N_{k,R}A_R \,\, , \qquad B_k=\sum_R N_{k,R}B_R}
where the sum is over the different messenger irreps, and
$N_{k,R}$ are the total Dynkin indices of the irrep $R$ with
respect to the gauge group $k$. Notice how the dependence on the
gauge group is trivial and factors out completely. The functions
$A_R$ and $B_R$ are universal in the sense that they depend only
on the mass parameters of the messengers with representation $R$.
In fact, they are identical to what one would compute for $n_R$
$U(1)$ messengers with charges $\pm 1$.

Since we are interested in models that are compatible with
unification, we should consider messengers in complete
representations of $SU(5)$.  The smallest $SU(5)$ representations
$\bf 5$ and $\bf 10$ can be decomposed under the usual matter
representations of the MSSM as \eqn\chargestotal{ {\bf \bar 5} = D
\oplus L\ , \qquad  {\bf 10} = Q\oplus U \oplus E.} So we will
restrict our attention to $R=Q,U,D,L,E$. Just for reference, the
Dynkin indices for these representations are \eqn\messch{\eqalign{
&N_{1,Q}={1\over10}, \ N_{1,U}={4\over5}, \ N_{1,E}={3\over5}, \
N_{1,D}={1\over5}, \ N_{1,L}={3\over10}\cr& N_{2,Q}={3\over2}, \
N_{2,L}={1\over2}\cr& N_{3,Q}=1, \ N_{3,U}={1\over2}, \
N_{3,D}={1\over2} }} where in the first line we have used the
standard GUT normalization for the $U(1)_Y$ charge.

The expressions \addcon\ immediately lead to a necessary
condition on the messenger content, in order for the model to
cover the full parameter space: we need messengers
transforming in at least three different irreps.  Otherwise, we do
not have three linearly independent functions $A_R$ and three
linearly independent functions $B_R$.

This means that any number
of messengers in ${\bf 5} \oplus {\bf \bar{5}}$ cannot cover the
parameter space (they have only two values of $R=D, L$). Next we can attempt to use messengers in a single copy of ${\bf
10} \oplus {\bf \bar{10}}$.  Here we have three values of $R=Q, U,
E$ and therefore three linearly independent constants.  However,
the result \MGMGbd\ in the $U(1)$ toy example discussion shows
that these constants are bounded,  $.37 <\kappa_R\equiv {A_R \over
| B_R|^2} <1$.  In particular, we cannot make the gauginos
arbitrarily heavy compared to the scalars.

As in the $U(1)$ example, we can avoid this difficulty by having
at least two copies of the representations and then using D-type
supersymmetry breaking. We are therefore led to the following
simplest possible models
 \eqn\possmod{ 2\times{\bf(10\oplus\bar{10}}) \qquad {\rm or} \qquad 2\times{\bf(5\oplus\bar5)\oplus10\oplus\bar{10}}\ .}
The latter is more ``minimal" since it has slightly smaller total
Dynkin index (and thus contributes slightly less to the MSSM gauge
coupling beta functions). However, the former is easier to
analyze, since we can now build a theory that is three copies of
the two-messenger models discussed in the previous section, one
for each irrep in the ${\bf 10}$.  The small SUSY breaking result
\coverresult\ is then enough to show that we can in fact cover the
parameter range.  This is true even if we take universal fermion
mass for each {$\bf 10\oplus\bar{10}$} factor, so we can cover the
parameter space without introducing supersymmetric GUT-breaking
splittings in the messenger sector. This shows that covering the
parameter space is compatible with unification, up to possible
threshold corrections coming from the SUSY-splittings.

The analysis of a theory with messenger content
$2\times{\bf(5\oplus\bar5)\oplus10\oplus\bar{10}}$ is slightly
different since the ${\bf 10\oplus\bar{10}}$ representations must
have pure F-type breaking. In particular, the $Q$, $U$, and $E$
type messengers must satisfy \MGMGbd\ and so \eqn\MGMten{
0.37<\kappa_R<1  \quad {\rm for}\,\,R=Q, U, E } Substituting
\MGMten\ into \addcon, we find six equations for seven non-compact
variables ($A(D)$, $A(L)$, $B(D)$, $B(L)$, $B(Q)$, $B(U)$, and
$B(E)$) and three compact variables ($\kappa_{Q,U,E}$). However,
it is not completely obvious that a real solution exists, because
the substitution is quadratic in $B(Q)$, $B(U)$ and $B(E)$. One
can check that this is always possible if we take
$\kappa_Q>\kappa_E,\kappa_U$. Note that this takes us outside the
small SUSY-breaking limit (where $\kappa=1$) for the $E$ and the
$U$ messengers.

These results show that there cannot be any additional field
theoretic restrictions on the GGM parameter space.  Another
consequence of this result is the following.  Assume that all the
soft terms are measured someday, and our two sum rules \massrell\
are satisfied.  Then, we can derive the six numbers $A_k$, $B_k$
and try to match them with a more microscopic theory.  Our result
here shows that whatever these numbers are, we'll be able to
obtain them from weakly coupled messengers.  In fact, we'll be
able to do it in more than one way.  This implies that the gaugino
and sfermion masses alone will not be enough to distinguish
between different gauge mediation scenarios. More input, such as
the messenger scale or the SUSY-breaking scale (equivalently, the
gravitino mass), will be needed in order to break this degeneracy.

\bigskip
\bigskip
\noindent {\bf Acknowledgments:}

We would like to thank N.~Arkani-Hamed, M.~Dine, Z.~Komargodski,
Y.~Shadmi, and Y.~Shirman for useful discussions. The work of MB was
supported in part by NSF grant PHY-0756966.  The work of PM and NS
was supported in part by DOE grant DE-FG02-90ER40542. The work of
DS was supported in part by NSF grant PHY-0503584.  Any opinions,
findings, and conclusions or recommendations expressed in this
material are those of the author(s) and do not necessarily reflect
the views of the National Science Foundation.

\appendix{A}{General results on the effective supertrace}
In this appendix we analyze the effect of integrating out massive
modes at tree-level in a renormalizable theory. In particular, we
will be interested in the supertrace over the spectrum of the
low-energy effective theory. We will assume that the low-energy
theory is described by a non-linear sigma model without gauge
interactions. Then the supertrace over the light modes is given by
the following general formula \refs{\GatesNR,\WessCP}:
\eqn\strformula{ {\rm STr}\CM^2=2R_{c\bar k}g^{\bar k a}g^{\bar b
c}W_{a}W^*_{\bar b} } where the indices run over the chiral
superfields $\Phi^a$ comprising the low-energy effective theory;
$g^{\bar a b}$ is the inverse K\"ahler metric; $R_{a\bar b}$ is
the Ricci tensor associated with the K\"ahler metric, and $W$ is
the effective superpotential.

We will show that integrating out massive chiral matter results in
a negative semi-definite Ricci tensor, so ${\rm STr}\CM^2\le 0$ in
this case. We then show that integrating out massive vector fields
results in an indefinite Ricci tensor and correspondingly a
supertrace of indefinite sign.

\subsec{Integrating out massive chiral matter}

Consider the most general renormalizable theory of heavy chiral
superfields $H^A$ coupled to light chiral superfields $\ell^a$.
This must have the form (we take the K\"ahler potential to be
canonical) \eqn\renormUV{
W={1\over2}\lambda_{Abc}H^A\ell^b\ell^c+{1\over2}M_{AB}H^AH^B+{1\over2}
m_{ab}\ell^a\ell^b+... } where the ellipsis contains unimportant
marginal and higher dimensional couplings, and $m\ll M$.
Integrating out the heavy fields yields the following equation of
motion \eqn\HEOM{
H^A=-{1\over2}(M^{-1})^{AB}\ell^T\lambda_{B}\ell+... }
Substituting this into \renormUV\ we obtain the effective
superpotential \eqn\Weff{ W_{\rm
eff}={1\over2}m_{ab}\ell^a\ell^b+\CO(\ell^4) } We also find the
following effective K\"ahler potential \eqn\Keff{ K_{\rm
eff}=\ell^\dagger\ell+{1\over4}\sum_A
\left|(M^{-1})^{AB}\ell^T\lambda_{B}\ell\right|^2... } It follows
that the Ricci tensor of the effective K\"ahler metric
\eqn\lightRicci{ R_{a\bar{b}}=-\partial_a(g^{\bar c d}g_{d\bar
b,\bar c})
 }
is at $\ell=0$ \eqn\leadinglambda{ R_{a\bar{b}}=-\delta^{\bar c
d}g_{a\bar b, \bar c d} =-\sum_A \left((M^{-1}\lambda)^{A}
(M^{-1}\lambda)^{\dagger \bar A}\right)_{a\bar b} } This is a sum
over negative semi-definite matrices, so it is also negative
semi-definite. It then follows from \strformula\ that the
effective supertrace over the light fields is non-positive. One
application of this result is to gauge mediation models of the
type discussed in section 3, where the $H^A$ fields are heavy
messengers and the $\ell^a$ are light messengers and SUSY breaking
fields.

\subsec{Integrating out massive vector superfields}

Next we consider what happens when one classically integrates out
massive vector superfields. Here it turns out that the Ricci
tensor of the effective K\"ahler metric is indefinite and
therefore the supertrace over the light spectrum is also of
indefinite sign.

The setup is as in \SeibergQJ; we will review it here. Consider a
gauge theory with matter chiral superfields $\Phi^a$ transforming
under gauge group $G$ (not necessarily simple), where
$a=1,\dots,N$ denotes the collective set of gauge and flavor
indices. Suppose that the $\Phi^a$ acquire supersymmetric vevs
$\phi_0$ which Higgs the entire gauge group. These vevs must lie
along the D-flat moduli space $\CM$ defined by the equations:
\eqn\Dflat{
 \phi_0^\dagger T^I \phi_0=0
 }
where $T^I$ are the generators of $G$. Now consider the
fluctuations around this point in moduli space: \eqn\Phifluct{
\Phi = \phi_0+\delta\Phi } We are interested in the effective
K\"ahler potential for these fluctuations induced by integrating
out the massive vector supermultiplets of $G$. In what follows we
will work in the unitary gauge discussed in \SeibergQJ
 \eqn\unitarygauge{
\phi_{0}^{\dagger}T^{I}\delta \Phi=0 } which guarantees that the
fluctuations lie within $\CM$. It will be
convenient to perform a unitary transformation so that
$\delta\Phi^{a=1,\dots,N-{\rm dim}G}$ satisfy \unitarygauge\ and
the other elements of $\delta\Phi$ are in the orthogonal subspace.

Now according to \SeibergQJ, the effective K\"ahler potential is
given by \eqn\Keff{ K_{\rm
eff}=\delta\Phi^{\dagger}\delta\Phi-{1\over2}(\delta\Phi^{\dagger}T^I\delta\Phi)
h_{IJ}^{-1}(\delta\Phi^{\dagger}T^J\delta\Phi)+\CO(\delta\Phi^6) }
where $h^{IJ}$ is the matrix \eqn\lambdadef{
h^{IJ}={1\over2}\Phi^{\dagger}\{T^I,T^J\}\Phi } (Note the analogy
with the previous subsection: $h_{IJ}^{-1}$ is analogous to
$M^{-1\dagger}M^{-1}$ and $T^I_{\bar b a}$ is analogous to
$\lambda_{Abc}$. The only difference is in the type of the
indices, which dictates how they are contracted.) As in the
previous subsection, we can compute the Ricci tensor at leading
order in the fluctuations. However, we must be careful not to
differentiate with respect to all the fluctuations $\delta\Phi^a$,
but only those which satisfy the gauge condition \unitarygauge. In
our convenient basis, these are simply the $a=1,\dots,N-{\rm
dim}\,G$ entries of $\delta\Phi^a$. So the metric is simply
\eqn\ginduced{ g_{a\bar b} = \delta_{a\bar b} -
(\delta\Phi^\dagger T^I)_a h_{IJ}^{-1} (T^J\delta\Phi)_{\bar b} -
(T^I)_{\bar b a}h_{IJ}^{-1}(\delta\Phi^\dagger
T^J\delta\Phi)+\CO(\delta\Phi^4) } with $a,b=1,\dots,N-{\rm
dim}\,G$. Therefore, the Ricci tensor at $\delta\Phi=0$ is:
\eqn\riccivector{ R_{a\bar b} =-\delta^{c\bar d}g_{a\bar b,c\bar
d} = (T^I)^c{}_{a}h_{IJ}^{-1}(T^J)_{\bar b c} +(T^I)_{\bar b a}
h_{IJ}^{-1}\Tr\,' T^J } Here the sum is only over indices in the
range $1,\dots,N-{\rm dim}\,G$, and $\Tr\,'$ refers to the
restricted trace over the subspace of fluctuations satisfying
\unitarygauge. Even though the full trace of $T^J$ must vanish due
to the anomaly condition, the restricted trace need not vanish
since the gauge symmetry is spontaneously broken. This is
important, because while the first term in \riccivector\ enjoys
definiteness properties, the second term obviously does not. Thus
there is no reason to expect the Ricci tensor to have any
definiteness property.  Indeed, it is straightforward to construct
simple examples where $R_{a\bar b}$ has both positive and negative
eigenvalues.\foot{For instance, consider a $U(1)$ gauge theory
with fields $\Phi^{1,2,3,4}$ having charges $q_1=+1$, $q_2=-1$,
$q_3=+q$ and $q_4=-q$ with $q\ne \pm 1$. The D-flat moduli space
is characterized by $\phi_0=(\Phi^1,\Phi^2,\Phi^3,\Phi^4)$ with
$\Phi^i$ satisfying the equation \eqn\dflatuiex{ |\Phi^1|^2
-|\Phi^2|^2+q(|\Phi^3|^2-|\Phi^4|^2) = 0 } Going to a point on
this moduli space, we can impose the gauge fixing condition
\unitarygauge\ by solving for $\delta\Phi^4$. Substituting back
into the K\"ahler potential \Keff\ gives the effective K\"ahler
potential for $\delta\Phi^{1,2,3}$. From this one can compute the
Ricci tensor at $\delta\Phi=0$ using $R_{a\bar b}=-\delta^{c\bar
d}g_{a\bar b,c\bar d}$. Then by varying $\phi_0$ and $q$ it is
easy to find places where $R_{a\bar b}$ has both positive and
negative eigenvalues.} Therefore we conclude in this case that the
effective supertrace can have either sign.

\appendix{B}{General results on multiple messenger models}

In this appendix we write down the GGM correlation functions for a
general messenger theory. We then explicitly show that a messenger
sector with non-vanishing supertrace generates contributions to
the scalar mass-squareds that are logarithmically divergent and
proportional to the supertrace.

As in section 4, let us restrict ourselves to the case that the
messengers are charged under a $U(1)$ gauge group with mass terms
\eqn\genmass{ V \supset
\xi_{ij}\phi_i\phi_j^*+\tilde\xi_{ij}\tilde\phi_i\tilde\phi_j^*+|M_i|^2
(\phi_i\phi_i^*+\tilde\phi_i\tilde\phi_i^*)+f_{ij}\phi_i\tilde\phi_j+
f_{ij}^*\phi_i^*\tilde\phi_j^*+M_{i}\psi_i\tilde\psi_i+
M_i^*\bar\psi_i\bar{\tilde\psi}_i } and $i=1,...,N$. Again, taking
the $\phi_i$ and $\tilde\phi_i$ to have $U(1)$ charge $+1$ and
$-1$ respectively, we find \eqn\currentsmmgm{\eqalign{
 & J(x) = \phi_i^*\phi_i-\tilde\phi_i^*\tilde\phi_i\cr
 & j_\alpha(x) = -\sqrt{2}i(\phi^*_i\psi_{i\alpha}-\tilde\phi_i^*\tilde\psi_{i\alpha})\cr
 & \bar j_{\dot\alpha}(x) = \sqrt{2}i(\phi_i\bar\psi_{i\dot\alpha}-\tilde\phi_i\bar{\tilde\psi}_{i\dot\alpha})\cr
 & j_\mu(x) = i(\phi_i\partial_\mu\phi_i^*-\phi_i^*\partial_\mu\phi_i-\tilde\phi_i
 \partial_\mu\tilde\phi^*_i+\tilde\phi^*_i\partial_\mu\tilde\phi_i)+
 \psi_i\sigma_\mu\psi_i-\tilde\psi_i\sigma_\mu\bar{\tilde\psi}_i
}} where we have implicitly summed over $i$.

Let us now write the various current two-point functions. To
perform the calculation, it will be convenient to change basis
from the gauge eigenstates appearing in \currentsmmgm\ to the mass
eigenstates via the following expressions \eqn\masseigenstates{
\phi_i=R_{ia}\cdot\varphi_a, \ \ \ \
\tilde\phi_i^*=R_{(i+N)a}\cdot\varphi_a } where $i=1,..., N$,
$a=1,..., 2N$, and $R$ is a $2N\times 2N$ unitary matrix. Let us
also denote the bosonic (fermionic) mass eigenvalues by $\mu_a$
($M_i$). Inserting \masseigenstates\ into \currentsmmgm, and
performing the contractions to evaluate the correlators, we find
\eqn\ccorrii{\eqalign{ &C_0(p) = \Big(\bar R_{ia}R_{ib}-\bar
R_{(j+N)a}R_{(j+N)b}\Big)\Big(\bar R_{kb}R_{ka}-\bar
R_{(l+N)b}R_{(l+N)a}\Big) I(p, \mu_a,\mu_b)\cr & C_{1/2}(p) ={p^2+\mu_a^2-M_i^2\over p^2}\Big(R_{ia}\bar R_{ia}+R_{(i+N)a}\bar
R_{(i+N)a} \Big)
 I(p,\mu_a,M_i) + {1\over p^2}\Big(J(\mu_a)-2J(M_i)\Big)\cr & C_1(p) =
{1\over3p^2}\Bigg((p^2+4\mu_a^2) I(p,\mu_a,\mu_a)+4(p^2-2M_i^2)
I(p,M_i, M_i)+4J(\mu_a)  -8J(M_i)+{\mu_a^2-2M_i^2\over4\pi^2} \Bigg) \cr
 &B=-4M_i R_{ia}\bar
R_{(i+N)a} I(0,M_i,\mu_a)
}}
where all indices are summed, and we
define \eqn\IJdef{\eqalign{ I(p,m_1,m_2)&=
\int{d^4q\over(2\pi)^4}{1\over((p+q)^2+m_1^2)(q^2+m_2^2)}\cr
 &={1\over16\pi^2}\Big(\log{\Lambda_q^2\over p^2}+1\Big)
 +{1\over16\pi^2p^2}\Big(m_1^2\log{m_1^2\over p^2}+m_2^2\log{m_2^2\over p^2}-m_1^2-m_2^2\Big)\cr
 &\qquad  +\CO\Big({1\over p^4},{\log p^2\over p^4}\Big)\cr J(m)&=\int{d^4q\over(2\pi)^4}{1\over q^2+m^2}={\Lambda_q^2\over16\pi^2}+{m^2\over16\pi^2}\log{m^2\over\Lambda_q^2}
}} where $\Lambda_q$ is a momentum cutoff for the $q$ integral. Simple consistency checks of the expressions in \ccorrii\ are the following.  As follows from supersymmetry, they all have the same asymptotic behavior, ${N\over 8\pi^2}\log {\Lambda^2\over p^2}$, at large $p$.  Also, since there are no massless particles in the loop, they are finite as $p\to 0$.\foot{We thank Thomas Dumitrescu for correcting some errors in the original version of \ccorrii.}

Let us now show that a non-vanishing messenger supertrace
necessarily generates a logarithmically divergent scalar
counterterm. Recall first the expression \Adef\ for the $A$
parameter \eqn\Adef{ A\equiv-\int{d^4p\over(2\pi)^4}{1\over
p^2}\Big(3C_1(p)-4C_{1/2}(p)+C_0(p)\Big) } Using \IJdef, \ccorrii\
and focussing on the $\CO(1/p^2)$ terms (one can check that the
$\CO(p^0,\log p)$ terms, and hence the dependence on $\Lambda_q$,
always vanish in \Adef), we find \eqn\strdiv{ \delta
A=-{1\over64\pi^4}\Big({\rm Tr}\mu^2-2{\rm
Tr}M^2\Big)\log\Lambda^2=-{1\over128\pi^4}{\rm
Str}\,\CM^2\cdot\log\Lambda^2 } where $\Lambda$ is the cutoff of
the $p$ integral in \Adef.

In this example we took the gauge group to be $U(1)$ and took all
the messengers to have charge $\pm 1$. More generally, one obtains
a charge-weighted supertrace, or to be precise \eqn\strdivii{
\delta A=-{1\over128\pi^4}\sum_R{\rm Str}\,
N_{R}\CM_R^2\cdot\log\Lambda^2 } where the supertrace is taken
over the subset of messengers transforming in irrep $R$ and $N_R$
is the Dynkin index of irrep $R$.

\listrefs

\end